\documentclass[authoryear,preprint,review,12pt]{elsarticle}

\usepackage{graphicx}
\usepackage{epsfig}

\usepackage{natbib}
\usepackage{amsmath, amsthm,amssymb}
\usepackage{lscape} 
\usepackage{epstopdf}
\usepackage{longtable}

\newcommand{\aap}{A\&AS }           
\newcommand{\aj}{AJ }           
\newcommand{\nat}{Nature }
\newcommand{\apj}{ApJ }
\newcommand{\icarus}{Icarus }

\newcommand{\apjl}{ApJL }

\newcommand{\araa}{ARA\&A }

\newcommand{\kq}{k_\textrm{P}}
\newcommand{\kw}{k_\textrm{S}}
\newcommand{\qq}{Q_\textrm{P}}

\newcommand{\mq}{m_\textrm{P}}
\newcommand{\mw}{m_\textrm{S}}
\newcommand{\Rq}{R_\textrm{P}}
\newcommand{\Rw}{R_\textrm{S}}
\newcommand{\Oq}{\Omega_\textrm{P}}
\newcommand{\Ow}{\Omega_\textrm{S}}
\newcommand{\dOq}{\dot{\Omega}_\textrm{P}}
\newcommand{\dOw}{\dot{\Omega}_\textrm{S}}
\newcommand{\aq}{\alpha_\textrm{P}}
\newcommand{\aw}{\alpha_\textrm{S}}

\journal{Icarus}

\begin{document}

\begin{frontmatter}

\title{The Mass, Orbit, and Tidal Evolution of the Quaoar-Weywot System}

\author[a,b]{Wesley C. Fraser \corref{cor1}}
\ead{wesley.fraser@nrc.ca}

\author[a,c]{Konstantin Batygin}
\author[a]{Michael E. Brown}
\author[d,e]{Antonin Bouchez}

\address[a]{Division of Geological and Planetary Sciences, MS150-21, California Institute of Technology, 1200 E. California Blvd. Pasadena, CA 91101 USA }
\address[b]{Herzberg Institute of Astrophysics, National Research Council, 5071 W. Saanich Road, Victoria, BC, V9E 2E7, Canada}
\address[c]{Institute for Theory and Computation, Harvard-Smithsonian Center for Astrophysics, 60 Garden St., Cambridge, MA 02138}
\address[d]{Giant Magellan Telescope Observatory, P.O. Box 90933, Pasadena, CA 91109}
\address[e]{Observatories of the Carnegie Institution, 813 Santa Barbara St. Pasadena, CA 91101}

\begin{abstract}
Here we present new adaptive optics observations of the Quaoar-Weywot system. With these new observations we determine an improved system orbit. Due to a 0.39 day alias that exists in available observations, four possible orbital solutions are available with periods of $\sim11.6$, $\sim12.0$, $\sim12.4$, and $\sim12.8$ days. From the possible orbital solutions, system masses of $1.3-1.5\pm0.1\times10^{21}$~kg are found. These observations provide an updated density for Quaoar of $2.7-5.0\mbox{ g cm$^{-3}$}$. In all cases, Weywot's orbit is eccentric, with possible values $\sim0.13-0.16$. We present a reanalysis of the tidal orbital evolution of the Quoaor-Weywot system. We have found that Weywot has probably evolved to a state of synchronous rotation, and have likely preserved their initial inclinations over the age of the Solar system. We find that for plausible values of the effective tidal dissipation factor tides produce a very slow evolution of Weywot's eccentricity and semi-major axis.  Accordingly, it appears that Weywot's eccentricity likely did not tidally evolve to its current value from an initially circular orbit. Rather, it seems that some other mechanism has raised its eccentricity post-formation, or Weywot formed with a non-negligible eccentricity.
\end{abstract}


\end{frontmatter}


\section{Introduction} \label{sec:intro}
Large Kuiper Belt objects with diameters $D\gtrsim1000$ km exhibit a broad range of densities, with values typically larger than $1.5\mbox{ g cm$^{-3}$}$ \citep{Buie1997,Rabinowitz2006,Brown2010,Fraser2010a}. In addition, many of the larger objects are found with small satellites only a few percent the size of the primary \citep{Brown2005b,Brown2007b,Fraser2010b} and that are icy in nature \citep{Barkume2006,Brown2006,Fraser2009b}. A natural explanation is one in which during the early phases of planetesimal growth, these large bodies accreted a large enough mass sufficiently rapidly to heat up and differentiate, producing silicate rich cores surrounded by icy mantles. Subsequent collisional evolution then stripped predominantly icy material, raising their densities above their primordial values. The range of densities exhibited by each object would then reflect the relative amounts of collisional bombardment that object suffered. Another bi-product of this process are the satellites of the larger objects  which are just collisional fragments from the mantle which were not ejected with a high enough velocity to escape, and remained bound to their primaries.

The large object Quaoar is an extreme case. Observations of its satellite, Weywot, presented by \citet{Fraser2010a} suggest it is the densest known KBO. The observed density of  $\rho=4.2 \pm1.3 \mbox{ g cm$^{-3}$}$ implies that Quaoar may consist almost entirely of silicate material surrounded by a very thin icy veneer \citep{Schaller2007b}. Such a result would imply that the amount of collisional bombardment suffered by Quaoar was significantly higher than that experienced by other large KBOs. 

The Quaoar-Weywot system presented a further peculiarity; Weywot appeared to be on an eccentric orbit \citep{Fraser2010a}. This was unexpected, as it seemed most likely that tidal evolution would circularize the orbits of the small satellites on short timescales consistent with the satellite of Eris, Dysnomia, which is found on a nearly circular orbit \citep{Brown2007b}. These two strange properties of the Quaoar-Weywot system warrant further investigation.

Here we present new observations of Quaoar and its satellite Weywot. In section~\ref{sec:observations} we present our observations of this binary system made with the Keck 2 telescope and the data reduction steps to identify Weywot within the images. In section~\ref{sec:orbit} we present a new determination of Weywot's orbit and colour, along with a more accurate determination of the Quaoar-Weywot system mass. In section~\ref{sec:tides} we present a re-analysis of the tidal evolution the Quaoar-Weywot and Eris-Dysnomia binaries. Specifically, we present eccentricity and semi-major axis evolution which considers tides raised on both bodies and an eccentric orbit of the satellite. In addition, we present some tidal evolution simulations of the Quaoar-Weywot system. We discuss possible evolution of the orbital parameters of Weywot and Dysnomia. Finally, we finish with a short discussion of the results in section~\ref{sec:discussion}.

\section{Observations and Data Reductions \label{sec:observations}}
Observations of the Quaoar-Weywot system were taken on June 7th, 2011 UT. During that night, Quaoar experienced an appulse with a 10.5" closest separation to a R=12.1 magnitude star. This allowed excellent adaptive optics image correction in natural guide star mode using the Keck 2 adaptive optics facility, resulting in image cores that had Full-Width at Half-Maxima of 55-66 milliarcseconds and a Strehl between 0.13 and 0.37, with 10-25\% of the light in the narrow core. Observations were taken with the Near Infrared Camera 2 (NIRC2) in the narrow camera mode resulting in a 0.01" pixel scale. Observations were taken in the K' filter using 15~s exposure times and a fixed sky position angle was maintained throughout the observations. The camera position angle was offset by 0.7 degrees from zero to account for a slight rotation that exists between the telescope FOV and the camera. A 3 point dither pattern was utilized to avoid the bad quadrant of NIRC2 and 15 images were taken at each dither point.

Images were reduced with standard techniques. Appropriate darks, biases, and dome flats were used to remove instrumental flat-field and bias patterns. Median stacks of the 15 images at each dither point were created from the de-biased and flat-fielded images. Background levels at a particular dither position and in the quadrant containing Quaoar were made by averaging the stacks of that quadrant before and after that particular dither. The result was a flat image with zero background to within the noise of the images.

\begin{figure}[hp] 
   \centering
   \includegraphics[width=5.5in]{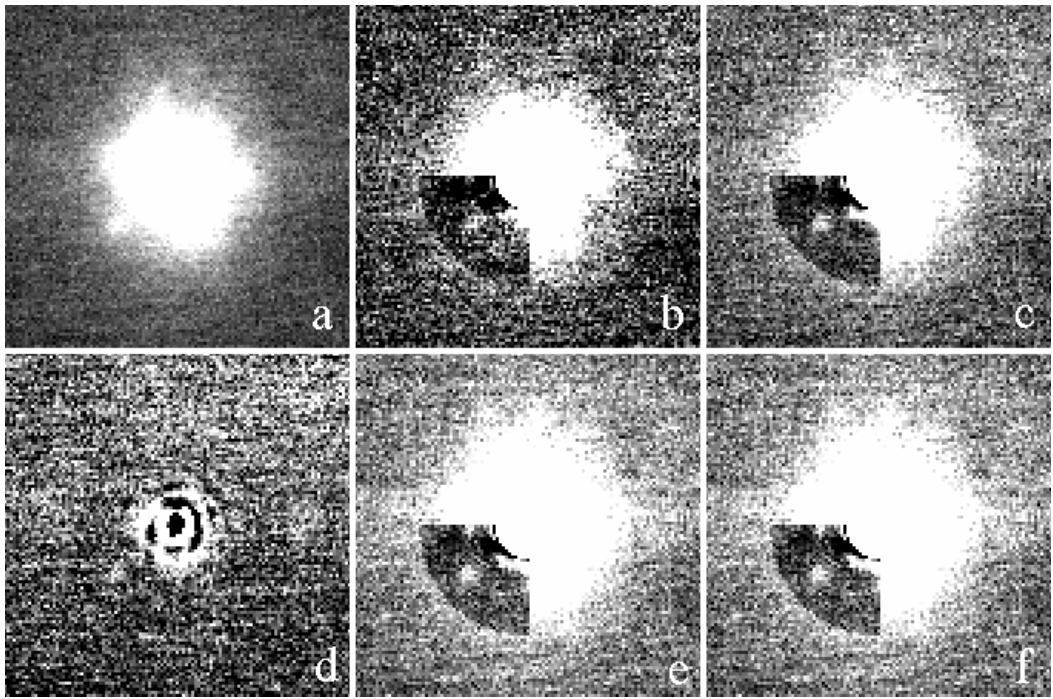}
   \caption{\textbf{a:} Median stack of all images centred on Quaoar. \textbf{b:} Median stack of first quarter of the images. The radial profile of Quaoar's image has been removed in the lower quadrant containing Weywot. \textbf{c:} As in \textbf{b}, but for the second quarter of the images. \textbf{d:} Stack of all images for which PSF subtraction was possible. \textbf{e:} As in \textbf{b}, but for the third quarter of the images. \textbf{f:} As in \textbf{b}, but for the fourth quarter of the images. In all images, North is up, East is to the left. \label{fig:stacks}}
\end{figure}

Identification of Weywot was made possible during a portion of the observations during which Quaoar passed close to a nearby star. Subtraction of Quaoar's point-spread function of sufficient quality for Weywot's easy identification was achieved when Quaoar was within 1.2" of the star, roughly 1/4 of the total of the time in which Quaoar was observed. In addition, the observations were taken in fixed sky position angle mode. As a result, the PSF and associated speckle pattern rotated by roughly $90^o$ demonstrating that the image of Weywot is real and not a PSF artifact. The image subtraction results along with a median stack of all images centred on Quaoar's position are shown in Figure~\ref{fig:stacks}. Weywot is easily seen South-East of Quaoar.

To determine accurate astrometric positions of Weywot with respect to Quaoar, 4 median stacks with equal equivalent exposure times were produced. For each stack, a radial profile of Quaoar's image was subtracted to reveal Weywot within the PSF wings. The radial profiles were generated using 15$^o$ wide radial slices separated by $15^o$ centred on Weywot's position. The resultant centroids in each image were not significantly altered by large variations in the radial slice parameters. The results are presented in Figure~\ref{fig:stacks}.

Astrometric positions were determined in three separate ways, PSF matching using the central 9 pixels of Quaoar's image, using Weywot's photo-centre, and gaussian profile fitting, all with a 5 pixel subsampling. The results of all three methods were quite consistent, with less than 1 pixel scatter. The final accepted positions are the average of all three methods, with a 1 pixel, or 0.01" uncertainty conservatively adopted to account for possible variations due to the centroid and radial profile fitting procedures. Weywot's motion can clearly be seen from stack to stack. These measurements are presented in Table~\ref{tab:positions}, along with available previous astrometric measurements.  Table~\ref{tab:positions} also includes corrections to the positions presented in \citet{Fraser2010a} which neglected the cosine of Quaoar's declination at time of the observations.

\section{Weywot and its Orbit \label{sec:orbit}}
From the four separate stacks, the Quaoar-Weywot system was found to have a K'-band flux ratio of $1:0.0034\pm0.0005$. Adopting similar albedos, Weywot has a size ratios of $\sim$5~\%, in rough agreement with previous observations \citep{Fraser2010a}

\begin{figure}[hp] 
   \centering
   \includegraphics[width=5.5in]{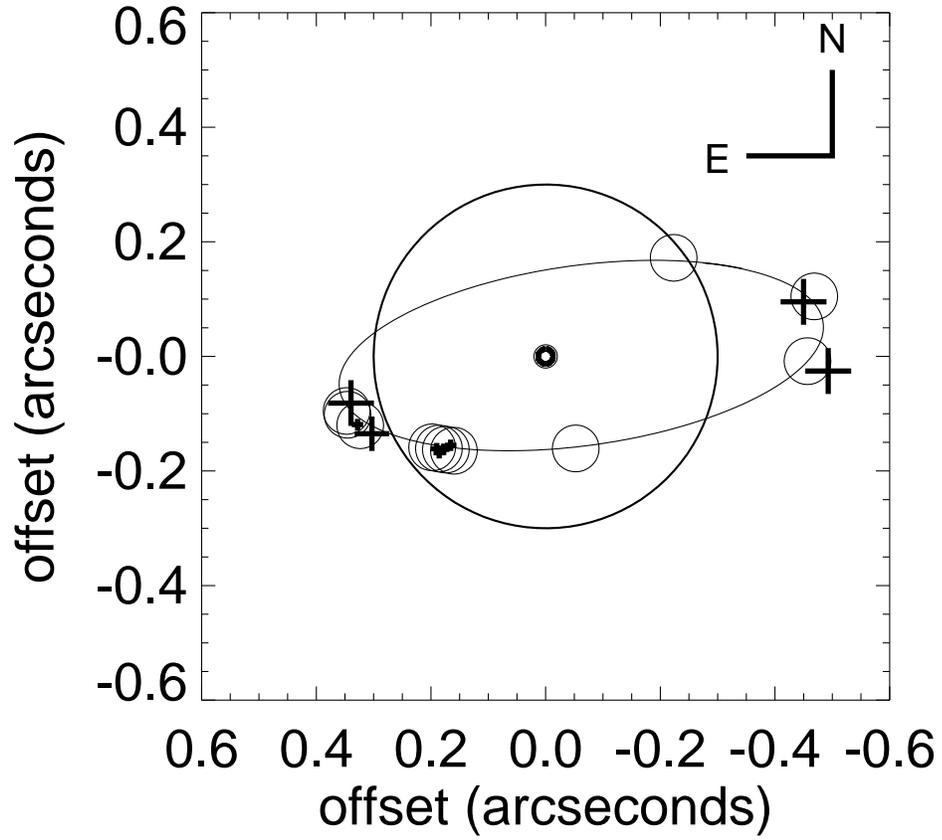}
   \caption{Weywot's observed offsets from Quaoar (shown at figure centre) and best-fitting orbit. Crosses are the observed position along with uncertainty at each epoch (see Table~\ref{tab:positions}). The best-fitting orbit (orbit 2 from Table~\ref{tab:orbits}) is shown as an ellipse, and the predicted positions at each observation from that orbit are shown as small circles. The large thick-lined circle represents the region in which Weywot would not have been detected in previous observations (see Fraser and Brown, 2010 for details). \label{fig:orbit}}
\end{figure}

Weywot's orbit was fit using the same methods as \citet{Fraser2010a}. That is, a maximum likelihood routine was used to consider all available detections and non-detections as a result of Weywot falling within Quaoar's image. The best-fit orbital solutions are presented in Figure~\ref{fig:orbit} and in Table~\ref{tab:orbits}. 

The observations do not have a sufficient temporal span to break the on-sky mirror degeneracy. In addition, the observations allow an alias in the period of 0.39 days, an alias that was missed by \citet{Fraser2010a} who only considered the best-fit orbital period of 12.4 days when determining the best-fit orbit. Formally, five orbital solutions corresponding to different local minima in the likelihood are found including that with the 12.4 day period were found when all data are considered. The compatibility of the different orbital solutions was tested by comparing the likelihood of each fit to the distribution of likelihoods found for that particular orbit. This was done by generating random simulated observations from one of the orbital solution. The simulated data were generated with gaussian noise with widths equal to the errors in the observed data. The simulated observations were fit and the process was repeated to generate a distribution of simulated likelihood values. For that fit, the probability $P(L_{ran}<L_{obs})$ of finding a random likelihood value smaller than the observed value, $L_{obs}$ was determined; values near 1 indicate a poor solution. From this test, only four periods of nearly $\sim11.68$, $\sim 12.04$, $\sim12.43$, and $\sim12.84$~days are possible solutions. Orbits with  $\sim11.3$~day periods are excluded at greater than the 99.8\% level.  The shortest acceptable period has $P(L_{ran}<L_{obs})>0.8$ and the longest has $P(L_{ran}<L_{obs})>0.95$. Thus, while formally consistent with the observations, the shortest and longest orbital periods are unlikely the correct ones.  Including their mirrors, eight acceptable orbits are found. The $12.04$~day prograde orbit provides the best solution. Additional observations are needed to determine which is the true orbit.

All acceptable orbits include non-zero eccentricities. A circular orbit with period $P=12.43$~days however, is possible with a moderately acceptable $L_{obs}=8.7$. To test the validity of a circular orbit over the best-fit eccentric orbit, we turn to the likelihood ratio test. We use the ratio to test the null hypothesis, that the improvement in the likelihood of an eccentric orbit over that of the circular orbit is insignificant. Assuming gaussian distributed measurement errors and in the limit of large sample size, the likelihood ratio $\chi=-2\log\frac{L_{obs}}{L_{obs,e=0}}$ is distributed as a chi-squared distribution with 1 degree of freedom. Here, $L_{obs}$ is the likelihood of the eccentric orbit fit, and $L_{obs,e=0}$ is the likelihood of the circular restricted fit. Of course, neither of the assumptions about the likelihood ratio necessarily hold true for our data, especially the former; 9 measurements and 2 non-detections certainly do not meet the requirement of a large sample size. As a result, we are forced to calibrate the likelihood ratio with the use of Monte-Carlo simulations. From the best-fit circular orbit we generate an artificial set of observations in the same manner as that used to determine $P(L_{ran}<L_{obs})$. Both an eccentric and a restricted circular orbit were fit to the simulated data, and the likelihood ratio, $\chi$ was determined. This process was repeated to generate a distribution of $\chi$ values. The advantage of this Monte-Carlo process is the complete avoidance of the assumption of a large sample size. All of our simulations include the same number of simulated observations as we actually have, and the resultant range of simulated likelihood ratios will reflect this.

In our simulations random likelihood ratio values larger than the observed value, $\chi=9.2$ were found in less than 1 in 1000 simulations. That is, the probability of the null hypothesis, that the eccentric fit results in an insignificant improvement over the circular fit, is less than 1 in 1000. This demonstrates that the eccentric orbit is a vast improvement over the circular fit. Similar improvements were even rarer for other periods and as such, we can formally exclude all circular orbit solutions at greater than the 3-$\sigma$ level.

It should be noted that the uncertainties on the astrometric positions are potentially overestimated. This can be seen from the best-fit orbit which has $P(L_{ran}<L_{obs})=0.14$; values near 0.5 are expected. This value suggests that the true uncertainty in the astrometric positions of Weywot, on average, may be as much as 25\% smaller than quoted. It should be made clear that this cannot be used to improve the orbital fits as it is virtually impossible to know which data points have over estimated uncertainties and which do not.

The four possible orbits (and their mirrors) provide a range of masses for the Quaoar-Weywot system of $1.3-1.4\times10^{21}$~kg. Given the best estimate of Quaoar's diameter of $890\pm70$~km \citep{Fraser2010a}, Quaoar's density must fall in the range $2.7-5.0\mbox{ g cm$^{-3}$}$.

\section{Satellite Orbital Evolution \label{sec:tides}}
The observations confirm that Weywot is on an eccentric orbit with eccentricity $e=0.13-0.16$. In addition, Weywot is on a highly inclined orbit, with an inclination of $\sim15^o$ ($152^o$ for the retrograde orbits). The fact that Weywot is on an eccentric orbit is surprising given the similar, albeit more massive, Eris-Dysnomia system which is found on a nearly circular orbit \citep{Brown2005b}. 

Here we revisit the question of tidal evolution. In \citet{Fraser2010a} we assumed (as is often done - see for example \citet{Noll2008}) that the dissipation from tides raised on the primary is negligible. We have found however, that this can be an important effect and here, we consider the mutual tides raised on both bodies in determining the orbital evolution of the system. 

To get a rough handle on whether or not it is possible that tidal effects can explain the differences in the binary orbits between the Quaoar-Weywot and Eris-Dysnomia systems, we first consider order of magnitude estimates of the tidal evolution timescales in an attempt to determine if reasonable tidal solutions exist that allow Weywot to maintain its eccentricity, or have it increased to its current value over the age of the Solar system. We choose to work within the weak friction tidal theory of \citet{Hut1981} in which the tidal potential is assumed to lag the direction of the tide raising body by a small time, $T$. Following \citet{Hut1981}, we place no restriction on the orbital eccentricity, $e$ and expand the equations to first order in inclination, $i$, which has the advantage of decoupling the $a-e$ dynamics from that of $i$. While little is known about the Weywot's inclination with respect to Quaoar's rotation pole, we found that the uncertainties in our estimates were largely dominated by errors in parameters other than inclination. 

As $e$ and $a$ are the most observationally constrained we consider these orbital parameters first. For the mutual tides raised on the binary system, the orbit-averaged rate of change of $a$ and $e$ are given by

\begin{equation}
<\dot{e}> = -27 \kq T_P \left(\frac{\mw}{\mq}\right) \left(\frac{\Rq}{a}\right)^5 \frac{e\,n^2}{\left(1-e^2\right)^{13/2}}\left[(1+D)f_3 -\frac{11}{18}(1-e^2)^{3/2}f_4\left(\frac{\Oq}{n}\right)\left(1+\frac{\Ow}{\Oq}D\right)\right]
\label{eq:EDot}
\end{equation}

\noindent 
and

\begin{eqnarray}
<\dot{a}> & = & -6 \kq T_P \left(\frac{\mw}{\mq}\right)\left(\frac{\Rq}{a}\right)^5 \frac{a \, n^2}{\left(1-e^2\right)^{15/2}}\left[(1+D)f_1 - \left(1-e^2\right)^{3/2}f_2 \left(\frac{\Oq}{n}\right)\left(1+\frac{\Ow}{\Oq}D\right)\right]
\label{eq:ADot}
\end{eqnarray}

\noindent
Here, $n$ is the mean-motion of the binary orbit. $m$, $R$, $T$, $k$ with appropriate subscripts where P refers to the primary and S refers to the secondary, are the masses, radii, tidal time lags, and tidal Love numbers of the two bodies. It is useful to relate the constant time-lag inherent to the tidal theory utilized here \citep{Hut1981} to the oft-quoted tidal quality factor, $Q$ \citep{Goldreich1966}. Following \citet{Efroimsky2007}, we express the relationship (assuming small lags) as $T = (2 Q |\Omega - n|)^{-1}$. Whenever quoting $Q$'s, we shall use the observed values of $\Omega_P$ and $n$, while taking pseudo-synchronous values for $\Omega_S$ since no observational information is available (note that this choice is only sensible if the pseudo-synchronization timescales for the satellites indeed turn out to be much shorter than the age of the solar system). The functions $f_1$, $f_2$, $f_3$,  $f_4$, and $f_5$ which are used below are functions of eccentricity which are of order unity for $e\ll1$ and are presented in closed form by \citet{Hut1981}. We have defined $D$ as

\begin{equation}
D=\frac{\kw}{\kq}\frac{T_S}{T_P}\left(\frac{\mq}{\mw}\right)^2\left(\frac{\Rw}{\Rq}\right)^5.
\label{eq:D}
\end{equation}

\noindent
Finally, $\eta=\aq\frac{\mq+\mw}{\mw}\left(\frac{\Rq}{a}\right)^2 (1-e^2)^{-1/2}\frac{\Oq}{n}$ is the ratio of the primary's rotational angular momentum to the orbital angular momentum  where $\alpha=\frac{I}{MR^2}$ with appropriate subscripts is the specific moment of inertia factor for each body. 

Our ultimate goal is to understand why Weywot is currently found on an eccentric orbit while Dysnomia is not.  Although it is presumed that Weywot and Dysnomia were formed through a collisional disruption event, the initial orbital and spin configurations of the satellites are poorly constrained. We therefore adopt the best-fit parameters to derive order of magnitude estimates of the tidal effects at current epoch. In particular, we choose the orbit with the highest maximum likelihood (orbit 2 in Table 2) for the Quaoar-Weywot system, and the orbital parameters found by \citet{Brown2005b} for the Eris-Dysnomia system. Because both possible rotation periods of Quaoar compatible with the light curve observed by \citet{Ortiz2003} greatly exceed the mean motion, here we shall use the shorter period of $8.84$ hr and keep in mind that the quoted estimates can be translated to the longer period by increasing $T_P$ by a factor of 2. For Eris, we adopt a $25.9$ hr rotation period as found by \citet{Roe2008}. 

Relevant uncertain (or unknown) parameters include the radii, rotation rates, masses, and moments of inertia, and the effective quality factors of both bodies. We estimate radii and masses of the satellites from the observed flux ratios of the primary-satellite pair, assuming equal albedos, and take the satellite densities to be $1 \mbox{ g cm$^{-3}$}$. The moments of inertia of the bodies are assumed to be that of uniform spheres, while the tidal Love numbers are given by  \citep{Peale1999}:

\begin{equation}
k = \frac{3/2}{1+\frac{19 \mu}{2 \rho g R}}
\label{eq:love}
\end{equation}

\noindent
where $g$ is the surface gravity. We adopt for the primary and satellite bodies rigidities of $\mu=4\times10^{11}$ and $\mu=4\times10^{10}\mbox{ dynes cm$^{-2}$}$ appropriate for rocky and icy bodies respectively. Note that the variations in density roughly cancel the variations in rigidity between the two materials, rendering $k$ an approximately unique function of radius. In all cases, adopting a wider range of these parameters does not change the main results significantly. Additional uncertain parameters will be discussed as needed. Despite the substantial number of unknown parameters, we can still determine instructive order of magnitude estimates of the tidal evolution of the binary systems at current epoch.

As already mentioned above, the current rotation rates of the satellites have not been observed. As will be shown below however, it is possible that the satellite's spins have evolved considerably over the age of the solar system due to tides. Thus, we adopt pseudo-synchronous angular velocities for the satellites to determine estimates of the variation timescales $\tau_{a}=\frac{a}{<\dot{a}>}$, $\tau_{e}=\frac{e}{<\dot{e}>}$. Here, it is important to note that if a satellite is sufficiently aspherical, in presence of considerable eccentricity it can become trapped in a spin-orbit resonance \citep{Goldreich1966}. This phenomenon may be of particular importance for Weywot. The associated librations of the satellite would provide an additional source of dissipation, which is not properly taken into account by the tidal theory of \citet{Hut1981}. In light of this, the degree of dissipation derived here may be viewed as an under-estimate for the secondaries.

\begin{figure}[hp] 
   \centering
   \includegraphics[width=5.5in]{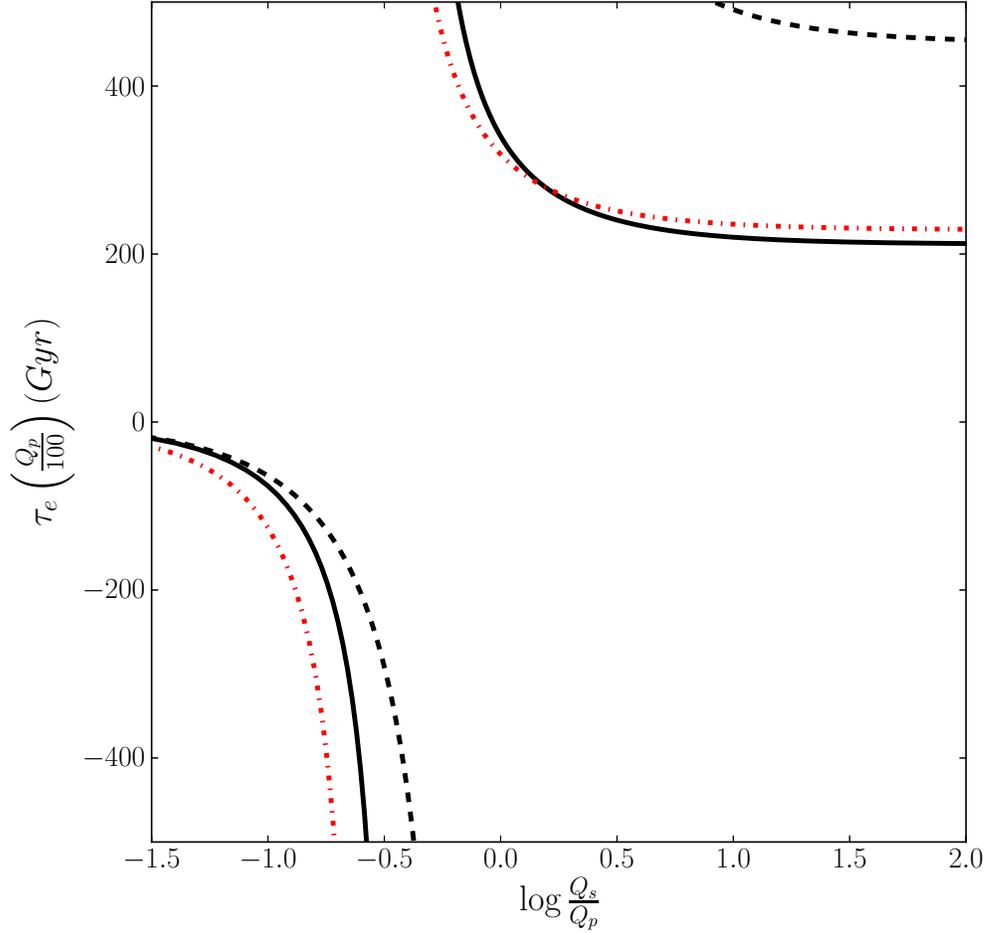}
   \caption{Eccentricity evolution timescales versus the ratio of tidal quality factor of the secondary and the primary bodies. The timescales are shown with  $\qq=100$ and scale linearly with $\qq$. Black solid and dashed lines correspond to the Quaoar-Weywot system with Quaoar rotating with 8.64 and 17.68 hr periods presented by \citet{Ortiz2003}. Red dash-dotted line shows the timescales for the Eris-Dysnomia system. Here we have adopted densities for the satellite of $1 \mbox{ g cm$^{-3}$}$ and a radius and albedo for Eris of $R_p=1170$~km and 0.96 (Sicardy, B. personal communication). \label{fig:tau_e}}
\end{figure}

The range of $\tau_{e}$ possible for both systems is presented in Figure~\ref{fig:tau_e} as a function of the tidal quality factor ratio of the secondary and the primary bodies. It should be noted that the range of timescales determined by our order of magnitude estimates are dominated by the uncertain tidal parameters, and the choice of any of the possible orbits for Weywot has little effect. 

From Figure~\ref{fig:tau_e} it might be interpreted that the order of magnitude estimate of the tidal evolution can qualitatively reproduce the contrasting eccentricities of Weywot and Dysnomia. That is, plausible values of $\frac{T_S}{T_P}$ can be chosen such that the $\dot{e}$ takes on positive values for Weywot, and negative values for Dysnomia. Analysis of Equation~\ref{eq:EDot} reveals that tidal eccentricity growth for Weywot requires $T_S/T_P \lesssim 40$ (corresponding to $Q_S/Q_P \gtrsim \frac{1}{2}$).

The eccentricity evolution timescales however, are much too long. Even if we set $T_S = 0$ (which corresponds to a completely non-dissipative secondary, and yields the fastest eccentricity growth), the growth timescale for Weywot is $\tau_\textrm{e} \simeq 6.3 \times Q_P$ Gyr and the damping timescale for Dysnomia is at least the age of the Solar System. These timescales can only be made considerably less than the age of the Solar system with unphysically small Q values, certainly incompatible with the values of $Q\sim100-500$ that have been inferred for other icy bodies such as the Galilean satellites \citep{Goldreich1966,Peale1986}. It is important however, to keep in mind that these estimates are made using current orbital parameters.


Using the above argument, we can quantify the sense of tidal evolution of the semi-major axis of the Weywot-Quaoar binary. Examination of Equation~\ref{eq:ADot} reveals that tidal decay of the orbit requires $T_S/T_P \gtrsim 330$ provided the current orbit. If we argue that Weywot's eccentricity has been monotonically increasing due to tidal effects over the system's lifetime, requiring $T_S/T_P \lesssim 40$, inward tidal migration can be ruled out. The fastest orbital growth however, (again setting $T_S = 0$) is characterized by a timescale $\tau_{\textrm{a}} \simeq 14.6 \times Q_P$ Gyr, much longer than the age of the Solar system.

We next turn our attention to the spin-states. The orbit-averaged rate of change of the spin velocities, $\Oq$ and $\Ow$ of primary and secondary bodies can be written as

\begin{equation}
<\dOq>=3 \kq T_P \left(\frac{\mw}{\mq}\right)^2\left(\frac{\Rq}{a}\right)^3\frac{n^3}{\left(1-e^2\right)^{6}}\frac{1}{\aq}\left[f_2-(1-e^2)^{3/2}f_5\frac{\Oq}{n}\right]
\label{eq:OmegaDotp}
\end{equation}

\noindent
and
\begin{equation}
<\dOw>=3 \kw T_S \left(\frac{\mq}{\mw}\right)\left(\frac{\Rw}{a}\right)^3\frac{n^3}{\left(1-e^2\right)^{6}}\frac{1}{\aw}\left[f_2-(1-e^2)^{3/2}f_5\frac{\Ow}{n}\right]
\label{eq:OmegaDots}
\end{equation}

\noindent
respectively. The construction of the tidal equations used here is such that the spins approach pseudo-synchronization at all times. On its current orbit, Weywot's pseudo-synchronization timescale approximately evaluates to $\tau_{\Omega_{\textrm{W}}} \simeq 2.4 \times 10^{-5} \times Q_S $ Gyr. Thus, even for the highest quality factors, densities, and radii that we can consider, the synchronization timescale for Weywot is no more than 100 Myr. A synchronization timescale roughly a factor of 50\% shorter is found for Dysnomia. As a result, it seems likely that  both Weywot and Dysnomia will have attained states of near synchronous rotation by the current epoch, although we again stress that in reality, Weywot's significant eccentricity may allow for capture into a high-order spin-orbit resonance.

Comparison of Equations~\ref{eq:OmegaDotp} and \ref{eq:OmegaDots} reveals that the synchronization timescales for the primaries is a factor of $\sim\left(\frac{\kw}{\kq}\right) \left(\frac{\mq}{\mw}\right)^3\left(\frac{\Rw}{\Rq}\right)^3\sim\left(\frac{\Rq}{\Rw}\right)^{6}\gg10^5$ longer than for the secondaries.Thus, the synchronization timescales for the primary bodies are $\gtrsim 100$ Gyr. As a consequence, the observed rotation rates of both Quaoar and Eris are likely to be very close to their primordial values.

Finally, we consider the binary orbital inclination. That is, the inclination, $i$, of the orbit with respect to the primary's spin axis. For small inclinations, the variation is well approximated by

\begin{equation}
<\dot{i}>=-3 \kq T_P \left(\frac{\mw}{\mq}\right) \left(\frac{\Rq}{a}\right)^5 \frac{ n^2 i}{\left(1-e^2\right)^{13/2}}\left[\frac{f_2}{\eta}-\frac{1}{2}\left(\frac{1}{\eta}-1\right)\frac{\Oq}{n}\left(1-e^2\right)^{3/2}f_5\right]
\label{eq:iDot}
\end{equation}

\noindent 
For Weywot near a state of synchronous rotation, the inclination damping timescale is $\tau_{i_{\textrm{Q}}}=\frac{i}{<\dot{i}>}{\sim 70 \times Q_P} \mbox{ Gyr}$ for Quaoar. A similar timescale is found for Eris. Even if the tidal damping factors are extraordinarily small, the inclination damping timescale still remains longer than $\sim700$~Gyr. Thus, while little constraint is available as to the actual values of $i$ for either system, it seems likely that both systems have preserved their primordial inclinations to the current day. It is noteworthy that the spin-axes of the secondaries align orders of magnitude faster than their primaries.

Our findings suggest that at present, the tidal evolution of the Weywot-Quaoar binary is generally slow. This may not have always been the case. Consequently, it is important to consider evolutionary paths of the system, with an eye towards determining the conditions under which Weywot can attain its eccentricity over a period of 4.5 Gyr. To address this, we consider tidal evolution simulations with the current observed state as a required end result of the simulations. Simulations were run with particular initial conditions searched to produce the required end state after 4.5 Gyr of integration. For these simulations, again we work within the context of the constant time-lag theory presented above and envision a giant impact formation scenario of the binary, similar to that inferred for the Moon. That is, we assume Weywot to have emerged from a proto-satellite disk on a near-circular orbit which then migrates outward achieving its current eccentricity 4.5 Gyr later.  

Taking Quaoar's rotation period to be 8.84 hrs, such an evolutionary sequence could not be found with reasonable values of $Q$. More precisely, for an initial eccentricity of $e_0 = 0.01$ and a dissipation factor $Q_\textrm{W}  = 100$, an evolution that produced the observed eccentricity requires unphysically small values of the dissipation factor for Quaoar,  $Q_{\textrm{Q}}\lesssim1$. Recall that physically, $Q=1$ implies complete dissipation of all the energy stored in a single libration cycle (it should also be noted that the weak-friction theory utilized in this work is strictly speaking inapplicable when $Q\lesssim10$.) Adjusting $Q_{\textrm{W}}$ to different values did not improve our results. No satisfactory result could be found with physically reasonable values of $Q_{\textrm{Q}}$.

The reader should be reminded that in our integrations, we assumed Weywot had an initial inclination sufficiently low such that $\cos i \sim1$. This assumption may be invalid; a non-zero initial inclination might be possible. The effect of a moderate inclination on the eccentricity evolution arises only indirectly through the change in semi-major axis (at second order). As a result, even  inclinations cannot change our results. It seems that simple tidal evolution is insufficient to drive Weywot's eccentricity to its current value over the age of the Solar system and it appears that tidal eccentricity excitation is not the dominant mechanism responsible for Weywot's orbit. The origins of this discrepancy could potentially be attributed to the simplicity of the tidal model we have utilized. In particular, we have ignored the possibly complex frequency and amplitude dependence of the tidal quality factors, a possibility that is difficult to test without a well-grounded theory of tidal dissipation \citep[see][for an excellent review]{Efroimsky2009}.

A plausible alternative scenario is one where Weywot's eccentricity is excited by some other mechanism and is then maintained due to a very long tidal circularization timescale. After all, as shown above, maintenance or even slow growth of eccentricity is naturally attained at the current orbit, given reasonable tidal quality factors. Possible mechanisms include collisions and close passages with other large bodies, or resonance passage during otherwise smooth tidal evolution. This would imply the presence of additional bodies in the system. It is further possible that Weywot did not coalesce out of a debris disk in orbit about Quaoar, but rather formed as a coherent collisional fragment that was ejected onto an orbit with non-zero eccentricity. The nature of such possibilities however, remains elusive.

\section{Discussion \label{sec:discussion}}

The observations presented here have significantly improved the accuracy of Weywot's orbital parameters given a particular orbital period and orientation. All available data however, are still unable to break the mirror degeneracy, or remove the aliasing which allows orbital solutions with four different periods. Regardless of the correct orbit, the observations have confirmed Weywot's significant orbital eccentricity, requiring an eccentricity larger than 0.1 to explain the observations. 

Analysis of the Quaoar-Weywot tidal evolution reveals that tidal dissipation in Quaoar is not necessarily negligible. It appears however, that tidal evolution is not the dominant mechanism responsible for Weywot's eccentricity. A plausible history is one in which Weywot primordially low eccentricity  was driven to one by some other mechanism, and subsequent tidal evolution had only small part in producing its current orbit. Such mechanisms might include collisions, or near passages with other massive bodies. Alternatively, it is possible that Weywot formed with a non-zero eccentricity, hinting that Weywot might be a coherent collisional fragment rather than a body which coalesced out of a disk of material in orbit about Quaoar.

We contrast this with the Eris-Dysnomia system which exhibits a similarly slow tidal evolution. This suggests that what ever mechanism produced Weywot's eccentricity eg. formation with high-$e$, was avoided by Dysnomia. This presents an intriguing diversity in small satellites around the large Kuiper Belt Objects. 

Our observations have confirmed Quaoar's large mass, which must be larger than $1.3\pm0.1\times10^{21}$~kg. Despite the remaining uncertainty in its mass, and the large uncertainty in its diameter, Quaoar still appears to be the densest known Kuiper belt object with a range of densities of $\rho=2.7-5.0\mbox{ g cm$^{-3}$}$.

The lower limit on Quaoar's density places it among the class of large, collisionally disrupted KBOs, along with objects like Haumea. These objects have apparently suffered significant post-differentiation, collisional evolution which has stripped away the majority of these object's icy content leaving only a thin veneer of ice left on their surfaces. In the case of Haumea, a single nearly catastrophic collision appears to be the culprit \citep[see for example][]{Leinhardt2010}. Quaoar's unusually high density suggests that it may be the result of a different collisional scenario than that which disrupted Haumea. On one hand, if a single collision is responsible, then Quaoar may have suffered a much more head-on impact than did Haumea, resulting in a much higher loss of its icy mantle. This may also explain Quaoar's nearly circular shape and slower spin - revealed by its rotational light-curve \citep{Ortiz2003} - as compared to the highly elongated and fast-spinning Haumea \citep{Rabinowitz2006,Lacerda2008}. Alternatively, Quaoar may have suffered a so-called hit-and-run impact in which Quaoar collided with a much larger body destroying Quaoar's icy mantle, but leaving its core intact \citep{Asphaug2006}. On the other hand, such a high density may be the result of multiple impacts which have slowly chipped away at Quaoar's primordial icy mantle.

\section{Acknowledgements}
W. F. would like to thank Alex Parker for his advice and help in determining correct orbital solutions. The data presented herein were obtained at the W. M. Keck Observatory, which is operated as a scientific partnership among the California Institute of Technology, the University of California, and the National Aeronautics and Space Administration.  The research upon which this paper is based was supported by National Aeronautic and Space Administration (NASA) Grant No. NNX09AB49G.

\newpage

\begin{longtable}{ccc}
   \caption{Quaoar and Satellite Positions. $\Delta$ RA and $\Delta$ DEC are the differences in position between the Quaoar and the satellite. The first 7 rows are from \citet{Fraser2010a}. \label{tab:positions}}\\
      \multicolumn{1}{c}{Epoch}  &\multicolumn{2}{c}{Satellite-Quaoar Offsets \footnotemark[1]} \\ 
      (JD+2453000) &  $\Delta$ R.A. (arcsec.) & $\Delta$ Dec. (arcsec.) \\
      \hline
     781.38031 & $0.328 \pm 0.01$ & $-0.119 \pm 0.01$ \\
   1179.12990 & $0.303 \pm 0.03$ & $-0.135 \pm 0.03$\\
   1535.70263 & $-0.49\pm 0.04$ & $-0.02\pm 0.04$ \\ 
   1540.56061 & $0.34\pm 0.04$ & $-0.08\pm 0.04$ \\
   1546.18353 & $-0.45\pm 0.04$ & $0.09\pm 0.04$\\
   1550.31485 & - & - \\
   1556.44075 & - & - \\
   2719.82856 & $0.17\pm 0.01$ & $-0.16\pm 0.01$\\
   2719.89737 & $0.18\pm 0.01$ & $-0.16\pm 0.01$\\
   2719.95760 & $0.19\pm 0.01$ & $-0.17\pm 0.01$\\
   2720.01759 & $0.20\pm 0.01$ & $-0.16\pm 0.01$\\
   \hline
   \footnotetext[1]{Values quoted in \citet{Fraser2010a} corrected to include $\cos{ \mbox{Dec}}$ term.}
\end{longtable}

\newpage

\scriptsize
\begin{longtable}{lllll}
   \caption{Orbital Solutions$^1$ \label{tab:orbits}} \\
       & Orbit 1 & Orbit 2 & Orbit 3 & Orbit 4 \\ \hline
    $\log(-L_{obs})$ 					& 7.40 				& 3.55 				& 4.1 				& 10.2 \\
    $P(L_{random}<L_{obs})$     		& 0.74 				& 0.14 				& 0.24 				& 0.04 \\
    Period (days)  					& $11.6836\pm0.0002$ 	& $12.0457\pm0.0002$  	& $12.4314\pm0.0002$ 	& $12.8428\pm0.0002$ \\
    Semi-major Axis ($10^4$ km) 		& $1.31\pm0.02$ 		& $1.35\pm0.02$ 		& $1.39\pm0.02$ 		& $1.45\pm0.02$ \\
    Eccentricity 						& $0.161\pm0.006$ 		& $0.152\pm0.005$ 		& $0.137\pm0.006$  	& $0.164\pm0.009$ \\
    Inclination (deg) 					& $14.8\pm0.7$ 		& $14.8\pm0.7$ 		& $15.8\pm0.7$ 		& $15.5\pm0.7$ \\	
    Longitude of Ascending Node (deg) 	& $15.5\pm0.8$ 		& $2.43\pm0.7$  		&  $1.0\pm0.7$ 		& $1.7\pm0.8$ \\
    Argument of Perihelion (deg) 		& $358.6\pm0.8$ 		& $8.44\pm0.7$  		& $335.0\pm0.7$ 		& $301.8\pm0.7$ \\
    Epoch of Perihelion -2454551 (JD)	& $2.67\pm0.1$		& $2.64\pm0.1$		& $1.58\pm0.1$ 		& $0.658\pm0.1$ \\
    System Mass ($10^{21}$ kg) 		& $1.30\pm0.09$		& $1.34\pm0.09$ 		& $1.39\pm0.09$ 		& $1.47\pm0.1$ \\
    Density (g cm$^{-3}$) \footnotemark[2] 	& $3.5 \pm0.8$ 		& $3.6 \pm 0.9$ 		& $3.8 \pm 0.9$ 		& $4.0\pm0.9$ \\
	\hline
    \multicolumn{4}{c}{Mirrors} \\
    	\hline
    $\log(-L_{obs})$ 					& 8.71  				& 3.53 				&  3.77				& 10.2 \\
    $P(L_{random}<L_{obs})$  			& 0.83 				& 0.13 				&  0.18 				& 0.04\\
    Period (days)  					& $11.6856\pm0.02$ 	& $12.0476\pm0.0002$ 	&  $12.4331\pm0.0002$ 	& $12.8443\pm0.0002$\\
    Semi-major Axis ($10^4$ km) 		& $1.30\pm0.02$ 		& $1.37\pm0.02$ 		& $1.41\pm0.02$ 		& $1.46\pm0.02$ \\
    Eccentricity 						& $0.154\pm0.006$ 		& $0.146\pm0.006$ 		& $0.128\pm0.005$ 		& $0.157\pm0.008$ \\
    Inclination (deg) 					& $151.2\pm0.7$ 		& $152.1\pm0.7$ 		& $150.7\pm0.7$ 		& $150.5\pm0.7$ \\
    Longitude of Ascending Node (deg) 	& $330.7\pm0.9$  		& $341.0\pm0.8$ 		& $341.3\pm0.8$ 		& $339.2\pm0.8$ \\
    Argument of Perihelion (deg) 		& $3.5\pm0.8$  		& $22.5\pm0.7$ 		& $344.1\pm0.7$ 		& $311.3\pm0.7$ \\
    Epoch of Perihelion -2454551 (JD)	& $2.5\pm0.1$			& $2.9\pm0.1$			& $2.7\pm0.1$ 			& $0.7\pm0.1$ \\
    System Mass ($10^{21}$ kg) 		&  $1.28\pm0.09$ 		& $1.41\pm0.09$  		& $1.43\pm0.1$ 		& $1.51\pm0.1$ \\
    Density (g cm$^{-3}$) \footnotemark[2] 	& $3.5 \pm 0.9$ 		& $3.8 \pm 0.9$ 		& $3.9 \pm 0.9$ 		& $4.1\pm0.9$ \\
    \hline
    \footnotetext[1]{Uncertainties are the $1-\sigma$ confidence limits on the relevant parameter with all others held at their best-fit values. Angles are with respect to the J2000 ecliptic.}
    \footnotetext[2]{Assuming a diameter for Quaoar of $D=890\pm70$ km.}
\end{longtable}
\normalsize

\newpage

\newpage

\newpage


\end{document}